\documentclass[multphys,vecphys]{svmult}
\usepackage{makeidx}     % allows index generation
\usepackage{graphicx}    % standard LaTeX graphics tool
\usepackage{multicol}    % used for the two-column index
\makeindex             % used for the subject index
\begin{document}

\title*{Explosion Models for Thermonuclear Supernovae Resulting from 
Different Ignition Conditions}
\titlerunning{Ignition Conditions and SNIa Models} 
\author{Eduardo Bravo\inst{1,2} \and Domingo Garc\'\i a-Senz\inst{1,2}}
\authorrunning{Eduardo Bravo \and Domingo Garc\'\i a-Senz}
\institute{Departament de F\'\i sica i Enginyeria Nuclear, Universitat
Polit\`ecnica de Catalunya, Av. Diagonal 647, Barcelona (Spain) \\
\texttt{eduardo.bravo@upc.es}
\texttt{domingo.garcia@upc.es}
\and Institut d'Estudis Espacials de Catalunya}
\maketitle

\abstract{We have explored in three dimensions 
the fate of a white dwarf of mass of $1.38 M_{\odot}$~ as a function 
of different initial locations of carbon ignition, with the aid of a SPH code. The calculated models cover 
  a variety of possibilities ranging from the simultaneous ignition of the
  central 
  volume of the star to the off-center ignition in multiple scattered spots. 
  In the former case, the possibility of a transition to a detonation when 
  the mean density of the nuclear flame decreases to $\rho\simeq 2~10^7$~  
  g.cm$^{-3}$~and its consequences are discussed. In the last case, the 
  dependence of the results as a function of the number of initial 
  igniting spots and the chance of some of these models to evolve to the 
  pulsating delayed detonation scenario are also outlined.}

\section{Statement of the Problem}
\label{sec:1}
\index{supernovae}
\index{multidimensional hydrodynamics}
\index{symmetry}

The analysis of light curves of many Type Ia Supernovae (SNIa) indicates that 
these events are not as homogeneous as one would desire to use them as perfect
standard candles. In fact, the inferred mass of Nickel ejected in the explosion,
estimated from the bolometric light curves \cite{sun}, could range within a
factor of ten. From a theoretical point of view, there remain a number of
fundamental issues to be solved before we can rely on the predictions of SNIa
modeling: pre-supernova evolution (progenitor systems, path of the white dwarf 
up to the Chandrasekhar mass, ignition conditions), physics of the explosion
(subgrid-scale physics vs. large-scale combustion, flame behavior at low
densities, deflagration--detonation transition), role of rotation and
magnetic fields, etc.

Although one--dimensional (spherical) models of SNIa have succeeded to explain a
wide range of observational properties, these kind of models unavoidably rely
on phenomenological descriptions of inherently multidimensional processes. Among them, 
flame acceleration due to hydrodynamic instabilities is of particular importance. Recent 
multi(3D)dimensional simulations of deflagrations in Chandrasekhar mass
white dwarfs have shown that the subsonic propagation of the burning front can
release enough energy to unbind the whole star and, maybe, can produce SNIa-like
explosions \cite{gam,rei}. Although the final test of these models will have to be done by
comparing self-consistent calculations of light curves and spectra (computed in
 three dimensions) with observations, one can already wonder if there is any
significant observational evidence of departure from spherical symmetry in SNIa.
In this respect, we note that:
1) Light curves are well described by a one-parameter family of curves
(e.g. \cite{phi}), 2) spectral absorption features by SiII show quite homogeneous profiles from
event to event \cite{tho}, and 3) polarization has not been detected in most 
SNIa, although there are a few exceptions \cite{kas}.
Supernova remnants (SNR) provide another means to constrain the geometry of the
explosions, although it depends also on the presence of
interstellar medium inhomogeneities and on the development of further
hydrodynamic instabilities. Still, some SNRs do not show large departures from
spherical symmetry (e.g. the blast wave of Tycho's SNR), which implies that both the supernova ejecta and the
interstellar medium possessed a high degree of symmetry in these cases (see also
the paper by Badenes et al in these proceedings). All these data point to
approximately spherical explosions in
which the chemical inhomogeneities are constrained to small-size clumps (a
quantitative criterion, proposed in \cite{tho}, is that the area of the 
individual clumps present at the photosphere has to be lower than $1-10\%$ of the 
photospheric area).

Our aim here is to address the following questions:
\begin{itemize}
\item Are 3D simulations of thermonuclear supernovae able to cover the full
range of light curve variations (up to a $10\times$ factor in $^{56}$Ni
production)?
\item How much the ejecta structure obtained in these simulations deviates from 
spherical symmetry (overall shape, small-scale clumping)?
\end{itemize}
We present the results from 3D hydrodynamic simulations starting from
different initial conditions. We have explored as well the possible outcomes
from deflagration--detonation transitions in 3D. It has to be stressed that all
calculations were carried out consistently, i.e. using the same
hydrocode \cite{gbs}, with the same numerical resolution, the same physics, and simulating
the whole white dwarf volume to avoid the introduction of artificial and
unrealistic symmetry conditions.

\section{Computed 3D models, $^{56}$Ni productivity and asymmetry}
\label{sec:2}
\index{deflagration}
\index{delayed detonation}

We have started from several ignition configurations, in which the carbon
runaway took place either in a central volume, slightly perturbed from 
spherical symmetry, or in multiple scattered spots or ''bubbles" \cite{gw95}.
In the second case, we have explored the sensitivity of the results to questions
like: How many bubbles ignite simultaneously? Which are the sizes of the
hot bubbles? Both theoretical analysis \cite{bg03} and 2D simulations 
\cite{hs02} do not seem to favor the formation of a large number of bubbles
igniting in phase (i.e. of the same size).
  
As for the deflagration--detonation transition, up to now there is no 
convincing mechanism that could account for
the acceleration of the burning front up to supersonic velocities, so we have
computed delayed detonation models making use of several different algorithms
for the initiation of the detonation. In one case, we started the detonation in
those points in which the fractal dimension of the flame was larger than 2.5,
while in the other cases, the condition for detonation ignition was that the
density were lower than $2\times10^7$~g~cm$^{-3}$ and the particles to be
detonated were selected according to their radius. The initial deflagration phase of all
the delayed detonation models was coincident with the deflagration model given in the first
row of Table \ref{tab:1}.

In Table \ref{tab:1}, there are listed the models computed and the main results
of the simulations. In all cases, the ignition started when the white dwarf
central density reached $1.8\times10^9$~g~cm$^{-3}$, and the simulations used
250,000 identical mass particles (each particle was about the same mass as
the Earth). The third column in Table \ref{tab:1} gives the final kinetic energy,
the fourth column gives the ejected mass of $^{56}$Ni, and the fifth column gives the 
ejected mass of unburned C and O.

\begin{table}[tb]
\centering
\caption{List of 3D simulations performed}
\label{tab:1}       
\begin{tabular}{llccc}
\hline\noalign{\smallskip}
Model category & Ignition configuration & $E_{\mathrm{kin}}$ &
$M(^{56}\mathrm{Ni})$ & $M(\mathrm{C+O})$ \\
 & & ($10^{51}$~erg) & ($M_{\odot}$) & ($M_{\odot}$) \\
\noalign{\smallskip}\hline\noalign{\smallskip}
Deflagration & Central volume & 0.26 & 0.27 & 0.65 \\
Deflagration & 6 equal size bubbles & - & 0.21 & - \\
Deflagration & 7 equal size bubbles & - & 0.21 & - \\
Deflagration & 10 equal size bubbles & 0.05 & 0.24 & 0.88 \\
Deflagration & 30 equal size bubbles & 0.44 & 0.56 & 0.58 \\
Deflagration & 90 random size bubbles & 0.45 & 0.58 & 0.58 \\
Delayed detonation & Transition where $D>2.5$ & 0.75 & 0.54 & 0.39 \\
Delayed detonation & Transition in central layers & 0.48 & 0.40 & 0.55 \\
Delayed detonation & Transition in medium layers & 0.51 & 0.43 & 0.51 \\
Delayed detonation & Transition in external layers & 0.33 & 0.32 & 0.65 \\
Pulsating & & & & \\
reverse detonation & 6 bubbles & 0.89 & 0.35 & 0.22 \\
\noalign{\smallskip}\hline
\end{tabular}
\end{table}

As can be seen from Table \ref{tab:1}, all the deflagration models
gave a kinetic energy too low to account for SNIa properties, with the exception of the two
simulations starting from a large number ($>30$) of bubbles. It is a
remarkable result that the explosion properties resulting from these two models
were almost the same, even though they started from very different initial conditions (in terms of number 
and sizes of hot bubbles). The results of these two
simulations were as well close to the ones given in \cite{rei}, for comparable
initial conditions, which reinforces the reliability of the outcome we
obtained. However, in the deflagration scenario there always remains a large mass of
unburned C and O, which would undoubtedly have to be detectable in the optical 
spectra of SNIa.

All our delayed detonation models gave healthy explosions with 
reasonable kinetic energies and a wide range of $^{56}$Ni masses. If nature 
provided a range
of conditions for transition to detonation as wide as we have explored, the delayed
detonation scenario would help to explain the diversity of SNIa
explosions observed. However, our 3D simulations have shown that
the delayed detonations are relatively inefficient in converting C and O to intermediate
mass elements or to iron peak elements (which is in sharp contrast with 1D models of
delayed detonations), which results in a lower velocity range than expected. The ultimate
reason for this behavior is the geometry of the distribution of fuel resulting from the 
previous deflagration stage. Due to the unavoidable distortion of the flame front caused by
hydrodynamic instabilities during subsonic burning, the initial conditions for delayed
detonation formation consist on large plumes of ashes coexisting at the same radii with
deep tongues of fuel (see Fig. 1). These ashes' plumes act as true barriers
obstructing the propagation
of the detonation waves and, in fact, decoupling large volumes occupied by C and O from
other regions which can hold propagating detonations. 

\begin{figure}[tb]
\centering
\includegraphics[height=3.8cm]{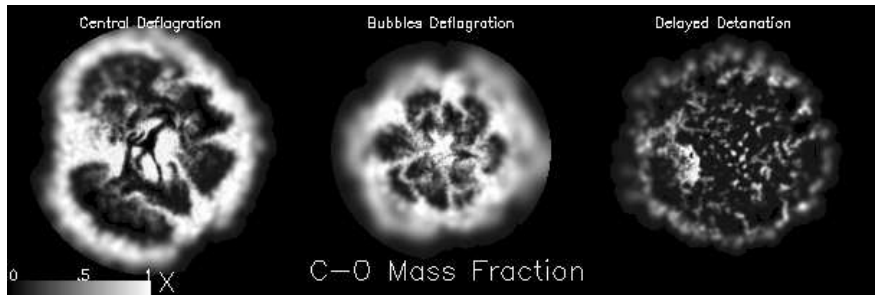}
%\picplace{5cm}{2cm} % Give the correct figure height and width in cm
%
\caption{Final distribution of unburned C-O in selected models. From left to right:
deflagration starting from a central volume, deflagration starting from 30 
bubbles, and delayed
detonation initiated where the flame fractal dimension was $D>2.5$}
\label{fig:1}
\end{figure}

\index{clumpy ejecta}
\index{flame propagation}

To discuss the geometrical properties of the explosions, we will refer to Fig. 
\ref{fig:1}, in which we show the distributions of the ejected C--O in
YZ slices. None of the models displays an overall
shape that departs largely from spherical symmetry, although the distribution of chemical
species presents irregularities of different sizes. However, the deflagration model starting
from a central volume (left image in Fig. \ref{fig:1}) is the most asymmetric.
 This is due to Rayleigh--Taylor
instabilities during flame propagation, which favor the formation of large--scale
structures that stand out in the final geometry of the ejecta. This configuration is in clear
disagreement with the limits to the size of clumps given in \cite{tho}. 
In the case of deflagrations starting from hot bubbles, the geometrical appearance of the
ejecta retains a larger degree of spherical symmetry, but still the chemical
inhomogeneities represent a large fraction of the radius of the star. We have estimated
that, in the model starting from 30 bubbles (center image in Fig. \ref{fig:1}), a typical
clump size is $\sim 10\%$ of the ejecta radius, still too large.

Delayed detonations (right image in Fig. \ref{fig:1}) are approximately
spherical in shape, and show only small clumps. We
have estimated a typical clump size of $\sim 4\%$ of the ejecta radius in these kind of
models. This could seem a striking result, as the delayed detonations were computed
starting from the final configuration of the deflagration shown in the left 
image in Fig.~\ref{fig:1}. The reason for this apparent contradiction is that the detonations were able to destroy the
large structures built up during the deflagration phase, thus allowing the formation of
only small-sized chemically differentiated clumps.

\section{Discussion and a new SNIa paradigm}
\label{sec:3}

To summarize, deflagrations computed in 3D produce large clumps, 
ejecta with low kinetic energy, reasonable quantities of $^{56}$Ni, and too much
unburned C--O ($>0.58~M_{\odot}$). On the other hand, multidimensional delayed detonations
produce smaller clumps, although they are not as efficient as their 1D counterparts at
rising the kinetic energy ($<0.75\times10^{51}$~erg) nor at burning C and O 
($M(C-O)>0.39~M_{\odot}$). In addition, we have seen that deflagrations
starting from a reasonable number of bubbles ($< 6-10$) fail to unbind the
white dwarf, which brings us to the following question: What is the fate of the
white dwarf following explosion failure?

\index{pulsating reverse detonation}
We have followed the evolution of the star during the first pulsation with the
3D SPH code. What came out from this multidimensional calculation was in fact
quite different from what is obtained in 1D pulsating models. Due
to the ability of bubbles to float to large radii in 3D, most of the thermal and
kinetic energy resided in the outer parts of the structure, which resulted in an
early stabilization of the central region (mostly made of C and O, i.e. fuel)
while the outer layers were still in expansion. Few
seconds later, an accretion shock formed at the border of the central nearly
hydrostatic core (whose mass was about $0.9~M_{\odot}$). Therefore, the
temperature at the border of the core increased to nearly $10^9$~K, on a
material
composed mainly by fuel but with a non-negligible amount of hot ashes, thus giving
rise to a highly explosive scenario. If a detonation were ignited at this point,
it would probably propagate all the way inwards through 
the core, burning most of it and producing an energetic 
explosion. We have called this new paradigm of SNIa
explosion mechanism the Pulsating Reverse Detonation. Currently, we are in the
process of completing the 3D calculation of its detonating phase. As a first 
evaluation of what one can expect from this new scenario, we have computed a 1D
model (last row in Table~\ref{tab:1}), which ended with the
largest kinetic energy, and the lowest mass of C and O from all the set
of models we have computed in 3D up to now. It turns out that it is worth
following the evolution of this scenario in 3D to see if it can derive in a new
reliable and competitive model of SNIa.

%%%%%%%%%%%%%%%%%%%%%%%%%%%%%%%%%%%%%%%%%%%%%%%%%%%%%%%%%%%%%%%%%%%%%%  }

%%%%%%%%%%%%%%%%%%%%%%%%%%%%%%%%%%%%%%%%%%%%%%%%%%%%%%%%%%%%%%%%%%%%%%

\printindex

\begin{thebibliography}{99.}

\bibitem{sun} N.B. Suntzeff: Optical, infrared, and bolometric light curves of
Type Ia supernovae. In: \textit{From Twilight to Highlight: The Physics of
Supernovae}, ed by W. Hillebrandt, B. Leibundgut (Springer, Berlin 2003) pp
183--192

\bibitem{gam} V.N. Gamezo, A.M. Khokhlov, E.S. Oran et al: Science \textbf{299},
77 (2003)

\bibitem{rei}  M. Reinecke, W. Hillebrandt, J.C. Niemeyer:  A\& A \textbf{391},
1167 (2002)

\bibitem{phi} M.M. Phillips, P. Lira, N.B. Suntzeff et al: AJ \textbf{118}, 
1766 (1999)

\bibitem{tho} R.C. Thomas, D. Kasen, D. Branch, D., E. Baron: ApJ \textbf{567}, 
1037 (2002)

\bibitem{kas} D. Kasen, P. Nugent, L. Wang et al: ApJ \textbf{593}, 788 (2003)

\bibitem{gbs} D. Garc\'\i a-Senz, E. Bravo, N. Serichol: ApJSS \textbf{115}, 
119 (1998)

\bibitem{gw95} D. Garc\'\i a-Senz, S.E. Woosley: ApJ \textbf{454}, 895 (1995)

\bibitem{bg03} E. Bravo, D. Garc\'\i a-Senz: Thermonuclear supernovae: Is
deflagration triggered by floating bubbles? In: \textit{From Twilight to 
Highlight: The Physics of Supernovae}, ed by W. Hillebrandt, B. Leibundgut 
(Springer, Berlin 2003) pp 165--168

\bibitem{hs02} P. H\"oflich, J. Stein: ApJ \textbf{568}, 779 (2002)

\end{thebibliography}
\end{document}